\documentclass[aps,twocolumn,showpacs,amsmath,amssymb]{revtex4}
\usepackage{graphicx}

\def\be{\begin{equation}}
\def\ee{\end{equation}}
\def\bea{\begin{eqnarray}}
\def\eea{\end{eqnarray}}

\def\a{\mbox{asin}\,}
\def\x{{\bf x}}
\def\v{{\bf v}}
\def\c{\mbox{conv}}
\newcommand{\w}[1]{(\ref{#1})}

\newcommand{\bra}[1]{\mbox{$\langle #1 |$}}
\newcommand{\ket}[1]{\mbox{$| #1 \rangle$}}

\begin{document}

\title{Necessary and sufficient condition for quantum-generated correlations}

\author{Ll. Masanes}
\affiliation{Dept. d'Estructura i Constituents de la Mat\`eria, Univ. Barcelona, 08028. Barcelona, Spain.}
\date{\today} 
\begin{abstract}
We present a non-linear inequality that completely characterizes the set of correlation functions obtained from bipartite quantum systems, for the case in which measurements on each subsystem can be chosen between two arbitrary dichotomic observables. This necessary and sufficient condition is the maximal strengthening of Cirel'son's bound.
\end{abstract}
\maketitle

The Principle of Causality imposes bounds on the correlations between space-like separated events: they have to emerge from interactions that took place in the past. Whether these interactions are classical or quantum makes a difference in this bounds.

It was Bell \cite{Bell} who first noticed that some correlations predicted by Quantum Theory (QT) are in contradiction with Local Variable Theories (LVTs), e.i. classical physics. This has been experimentally proven up to some loopholes \cite{Aspect}. A lot of work has been made to characterize the bounds of LVTs' correlations \cite{Fine,Werner,Zukowski,Collins}. These bounds are called Bell inequalities and are a nice frame to experimentally invalidate classical physics because, ideally, no assumptions and models for the experiments are necessary, just measuring correlations.

Little is known about the bounds for the correlations obtainable within QT: a necessary condition found by Cirel'son \cite{Cirelson}, some numerical results \cite{Filipp}, some general but partial results \cite{conv}, and a characterization in terms of a convex hull \cite{Werner}. This last description is useful for many purposes but in some situations it would be better to know the analytic shape of the bounds. This paper contains a complete and simple characterization of the bounds for the correlations attainable within QT for a scenario that is stated below. In the same foot as Bell inequalities these bounds provide a good frame to experimentally search for the existence of {\em superquantum} correlations \cite{Popescu}, without assumptions and models for the experiments. 

From a more practical point of view, the sharing of correlations among several parties is a useful resource for tasks like distributed computation \cite{cc} and secret key agreement \cite{cryptography}. Then, it is important to know if a given set of correlations is achievable with shared classical randomness, quantum entanglement, or it requires some amount of communication. One could say that Bell inequalities tell us impossibilities for Classical Information Theory. In the same way, the result of this work shows limitations for Quantum Information Theory.

\bigskip
The scenario that is considered in this work consists of two separated parties ---Alice and Bob--- sharing a bipartite system. Alice (Bob) can carry out two possible measurements, $A_0$ and $A_1$ ($B_0$ and $B_1$), with outcomes $1$ and $-1$. No assumption is made on the kind of systems they have. Then, although all observables are dichotomic we do not restrict to local two-dimensional systems. At space-like separated events Alice and Bob choose and perform one measurement each. With the observed results they can construct the vector of correlation functions:
\be
  {\bf x} = \begin{pmatrix} \langle A_0 B_0\rangle &\, \langle A_0 B_1\rangle &\, \langle A_1 B_0\rangle &\, \langle A_1 B_1\rangle \end{pmatrix}\ .
\label{correlations} \ee
The set of all possible vectors of correlators is a four-dimensional cube characterized by the eight trivial inequalities:
\be
  -1 \leq x_k \leq 1 \hspace{1cm} k=1,2,3,4
\label{trivial} \ee
What regions of this set are accessible depends on the theory that is used to describe the state of the compound system as well as the process of measuring it. In what follows two theories are considered.

\bigskip
{\em Local Variable Theories:} It was found in \cite{Fine,Werner} that a vector of correlation functions ${\bf x}$ can be obtained within a LVT if and only if it satisfies the eight inequalities:
\bea
  -2 \leq-x_1 + x_2 + x_3 + x_4 \leq 2 \nonumber \\
  -2 \leq x_1 - x_2 + x_3 + x_4 \leq 2 \nonumber \\
  -2 \leq x_1 + x_2 - x_3 + x_4 \leq 2 \nonumber \\
  -2 \leq x_1 + x_2 + x_3 - x_4 \leq 2 \label{chsh} 
\eea
We denote this set by $C$. Because $C$ is defined with linear inequalities it is a convex polytope, namely a four-dimensional octahedron \cite{Werner}. These eight inequalities are equivalent in the sense that, each one can be transformed into the other by interchanging parties, observables and outcomes. A representative of them is the well known CHSH \cite{CHSH}.

\bigskip
{\em Quantum Theory} gives the same predictions than LVTs when considering separable states, but in general it goes beyond. Cirel'son proved that the inequalities \w{chsh} cannot be violated by QT with a value larger than $2\sqrt{2}$ \cite{Cirelson}, for example
\be
  x_1 + x_2 + x_3 - x_4 \leq 2\sqrt{2}\ .
\label{cir} \ee
The main result of this work corresponds to the following theorem:

{\bf Theorem:} {\em A vector of correlation functions ${\bf x}$ is obtainable within QT if and only if it satisfies the conditions:}
\bea
  -\pi \leq- \a x_1 + \a x_2 + \a x_3 + \a x_4 \leq \pi \nonumber \\
  -\pi \leq  \a x_1 - \a x_2 + \a x_3 + \a x_4 \leq \pi \nonumber \\
  -\pi \leq  \a x_1 + \a x_2 - \a x_3 + \a x_4 \leq \pi \nonumber \\
  -\pi \leq  \a x_1 + \a x_2 + \a x_3 - \a x_4 \leq \pi \label{nonlinear} 
\eea
{\em where} $\a x$ {\em is the inverse of the sinus function.} The set of points fulfilling (\ref{trivial}) and (\ref{nonlinear}) is denoted by $Q$. In analogy to $C$, these eight inequalities are equivalent. 

\bigskip
{\bf Symmetries of $Q$ and s-order.} In what follows it is seen that the sets $C$ and $Q$ have two symmetries that simplify their study. Any permutation of the coordinates $x_i$ leave the sets of equations \w{trivial}, \w{chsh} and \w{nonlinear} unaltered. The same happens when changing the sign of two coordinates. This implies that the property of belonging to $C$ or $Q$ remains unchanged after applying these transformations. Notice that any point can be transformed into one satisfying 
\be
  x_1 \geq x_2 \geq x_3 \geq |x_4|\ . \label{J}
\ee
Then, it is enough to consider such points. Following the same notation as in \cite{ibm}, a vector is said to be s-ordered if it satisfies \w{J}. From now on, unless explicitly mentioned, any vector $\x$ is supposed to be s-ordered. Almost all the inequalities in \w{trivial}, \w{chsh} and \w{nonlinear} are automatically satisfied by these vectors. After carefully rejecting the useless inequalities we conclude that, for any s-ordered $\x$

${\bf x} \in C:$
\bea
  x_1\leq 1 \label{c1} \\ x_1+x_2+x_3-x_4 \leq 2 \label{c2} 
\eea

${\bf x} \in Q:$
\bea
  x_1\leq 1 \label{q1} \\ \a x_1 + \a x_2 + \a x_3 - \a x_4 \leq \pi \label{q2}
\eea

In what follows, we proceed on proving the Theorem. We have divided the proof into an initial declaration of simple facts, six lemmas, and a final reasoning (Proof of the Theorem). The reader not interested in technical details can skip all this part.


\bigskip
Let us start by defining $F$ as the set of points for which at least one of the inequalities in \w{nonlinear} is saturated. All s-ordered points in $F$ saturate inequality \w{q2} and satisfy the rest \w{nonlinear}. Then by symmetry, all points in $F$ satisfy \w{nonlinear}, which implies $F \subseteq Q$. Because for all correlation vectors \w{trivial} the conditions \w{nonlinear} are well-defined and continuous, the boundary of $Q$ ($\partial Q$) is the union of $F$ and points saturating at least one of the inequalities in \w{trivial}. 

\bigskip
{\bf Lemma 1:} $Q$ is a convex set.

In order to prove this statement let us consider s-ordered points belonging to the hypersurface $\partial Q$. The set of points saturating \w{q1},
\be
  x_1 = 1\ ,
\label{hp} \ee
 is a hyperplane, therefore it is convex. The set of points saturating \w{q2} are such that
\be 
  f(\x)=\pi \ ,
\label{surf} \ee 
where $f(\x)= \a x_1 +\a x_2 +\a x_3 -\a x_4$. To prove the convexity of this surface the gradient and the hessian matrix of $f(\x)$ have to be computed:
\bea
  && \nabla_i f(\x) = \frac{1}{\cos \gamma_i} \label{grad} \\
  && H(\x)_{ij} = \frac{\partial^2 f(\x)}{\partial x_i \partial x_j}= \frac{\sin \gamma_i}{\cos^3 \gamma_i} \delta_{ij}
\eea
where $\gamma_1=\a x_1$, $\gamma_2=\a x_2$, $\gamma_3=\a x_3$ and $\gamma_4=-\a x_4$. It is well known that if for each point $\x$ satisfying \w{surf} the matrix $H(\x)$ is positive in the subspace orthogonal to $\nabla f(\x)$, the surface \w{surf} is convex. Equation \w{surf} can be written as
\be
  \gamma_1 +\gamma_2 +\gamma_3 +\gamma_4 = \pi 
\label{eq} \ee
where each $\gamma_i$ belongs to the interval $[-\pi/2,\pi/2]$, this implies that at most one eigenvalue of $H(\x)$ is negative. When all the eigenvalues are positive there is no problem. Let us suppose without loss of generality that the eigenvalue corresponding to the eigenvector $\v_4=\begin{pmatrix}0&0&0&1\end{pmatrix}$ is negative. The vector belonging to the subspace orthogonal to the gradient which has maximal overlap with $\v_4$ is
\be
  \v_4 - \frac{\v_4\cdot\nabla f(\x)}{(\nabla f(\x))^2}\, \nabla f(\x)\ .
\ee
The expected value of $H(\x)$ with this vector is proportional (with positive constant) to
\be
  \sum_{i=1}^3 \frac{\sin \gamma_i}{\cos^5 \gamma_i} -\tan(\gamma_1+\gamma_2+\gamma_3)\left( \sum_{i=1}^3 \frac{1}{\cos^2 \gamma_i} \right)^2
\ee
where we have used \w{eq} to substitute $\gamma_4$, and therefore the sum $\gamma_1+\gamma_2+\gamma_3$ must belongs to $[\pi/2,3\pi/2]$. We have checked numerically that this expression is positive for the allowed $\gamma$'s. Now, we have seen that the surfaces \w{hp} and \w{surf} are convex. The fact that the unitary vectors orthogonal to \w{hp} and \w{surf} are equal in the intersection of both surfaces, together with the symmetries, warrant the convexity of $Q$. To see this fact let us take the limit $x_1\rightarrow 1$ ($\gamma_1 \rightarrow \pi/2$) in the vector \w{grad}. This limit is proportional to $(1\,0\,0\,0)$, which is the vector normal to \w{hp}. The symmetries imply that the unitary vector normal to $\partial Q$ is continuous everywhere. $\Box$ 

\bigskip
{\bf Lemma 2:} $Q$ is contained in the convex hull of the points saturating the inequalities \w{nonlinear}: $Q\subseteq \c F$.

\begin{figure}
 \includegraphics[width=3 cm]{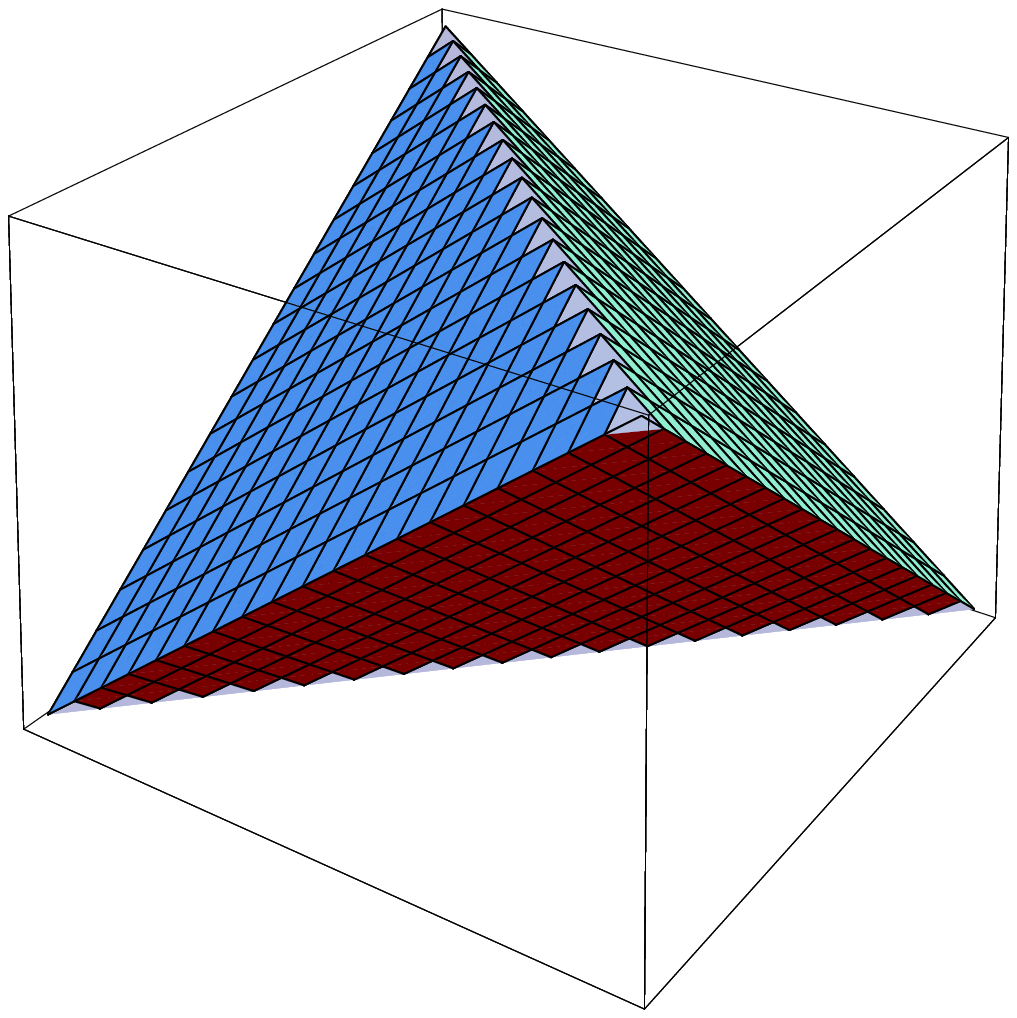}
 \hspace{.6cm}
 \includegraphics[width=3 cm]{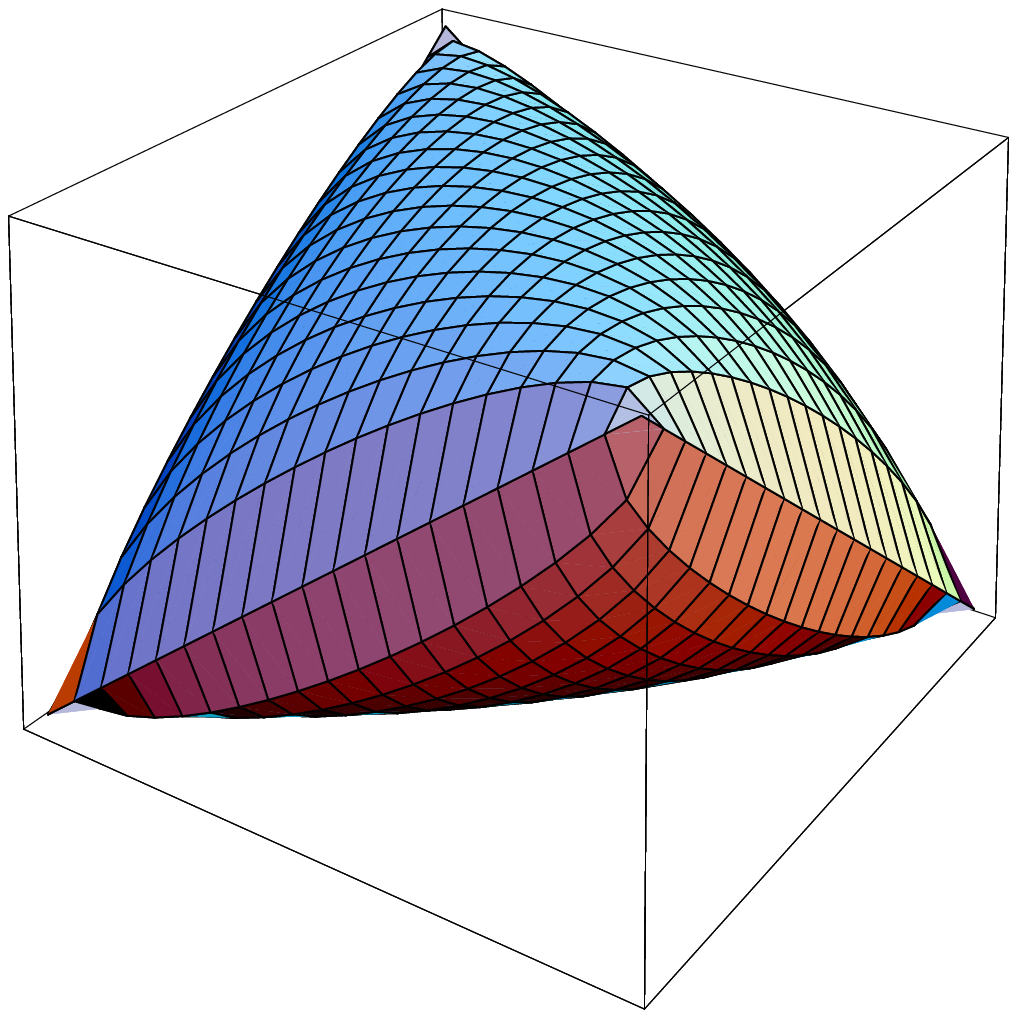}
 \caption{These figures are respectively the intersection of $C$ and $Q$ with the set of points satisfying $x_1=1$.}\label{fig}
\end{figure}

It is easy to see that any compact set belongs to the convex hull of its boundary \cite{lulu}, in our case $Q \subseteq \c \partial Q$. Using this we can prove the lemma by showing that $\partial Q$ is contained in the convex hull of $F$: $\partial Q \subseteq \c F$, because this implies that $\c \partial Q \subseteq \c F$, and therefore $Q\subseteq \c F$, which completes the proof. When studying the boundary of $Q$ we have seen that, all points in $\partial Q$ are in $F$ or saturate at least one inequality in \w{trivial}. In what follows it is seen that all points in $\partial Q$ which are not in $F$ belong to $\c F$, as we want to show. The points in $Q$ for which $x_1=1$ are the ones satisfying
\bea
  &x_4& \geq -\cos(\arcsin x_2 +\arcsin x_3 ) \label{p1} \\ 
  &x_4& \leq  \cos(\arcsin x_2 -\arcsin x_3 ) \label{p2}\ .
\eea
These conditions are obtained imposing $x_1=1$ in \w{nonlinear}, therefore, points saturating at least one of these inequalities belong to $F$. This set is plotted in right part of FIG. \ref{fig}. When $x_2$ or $x_3$ is equal to $\pm 1$ the right hand side of \w{p1} and \w{p2} coincide, which implies that the boundary of this set is made of points belonging to $F$. Using \cite{lulu} again it can be said that this set is contained in $\c F$, and by symmetry, the whole $\partial Q$ belongs to $\c F$, which finishes this proof. $\Box$ 

\bigskip
{\bf Lemma 3:} The set of all correlation vectors obtainable within QT is convex.

This result was proven in \cite{Werner,conv}. Nevertheless here we attach the proof for the sake of completeness. First of all, recall that any measurement can be written as a projective one by enlarging sufficiently the Hilbert space where it acts. Then, in the proofs of Lemmas 3 and 4 we only consider projective measurements. Now, our aim is to show that if ${\bf x}$ and ${\bf x}'$ can be obtained within QT, for any value of $\lambda$ in the range $[0,1]$, the point $\lambda {\bf x} + (1-\lambda) {\bf x}'$ is also a quantum correlation vector. Let us suppose that there exists a pair of observables for Alice ($A_a$ with $a=0,1$), a pair of observables for Bob ($B_b$ with $b=0,1$), and a density matrix $\rho$, for which $\mbox{tr}(A_a \otimes B_b \ \rho)$ yields the components of ${\bf x}$, and the same for ${\bf x}'$. To obtain the components of $\lambda {\bf x} + (1-\lambda) {\bf x}'$ we can construct the pair of observables for Alice $A_a \oplus A'_a$ (with $a=0,1$), the pair of observables for Bob $B_b \oplus B'_b$ (with $b=0,1$), and the density matrix $\lambda \rho \oplus (1-\lambda) \rho'$ which accomplish our purpose. $\Box$

\bigskip
{\bf Lemma 4:} The set of all correlation vectors obtainable within QT is $\c G$, where $G$ is the set of vectors 
\be
\begin{pmatrix} \sin\phi_1 &\, \sin\phi_2 &\, \sin\phi_3 &\, -\sin(\phi_1+\phi_2+\phi_3) \end{pmatrix}, 
\label{generators} \ee
for all values of $\phi_1,\phi_2,\phi_3$.

This result is the characterization in terms of a convex hull given in \cite{Werner}. The prove that we give here is simpler than the original one. To proof this lemma we proceed in two steps. First, we show that all quantum correlation vectors belong to $\c G$. Second, for each value of $\phi_1,\phi_2,\phi_3$, we explicitly provide the quantum states and the observables giving the correlations \w{generators}. This together with Lemma 3 implies that, $\c G$ is contained in the set of quantum correlation vectors. These two steps are sufficient to finish the proof.
First step: because the eigenvalues of $A_a$ are $\pm 1$ and adopting the convention that any operator exponentiated to zero gives the identity, a generic pair of observables for Alice can be written as $A_a = (A_1 A_0)^a A_0$. Recalling that any $A_a$ is unitary we have $A_a = P_0^A A_0 + e^{ia\alpha} P_1^A A_0$, with $P_0^A$ and $P_1^A$ being two projectors such that $P_0^A + P_1^A$ is the identity operator on Alice' Hilbert space. The same can be done for Bob's pair of observables $B_b$. The most general correlation vector predictable with QT is   
\be
  \langle A_a B_b \rangle = \sum_{r,s=0,1}  e^{i (ra\alpha + sb\beta)}\ \mbox{tr}\! \left[ \rho\, (P_r^A A_0) \!\otimes\! (P_s^B B_0) \right]
\label{expected} \ee
where $\rho$ is a unit-trace positive matrix acting on the tensor product of Alice' and Bob's Hilbert spaces. Let us put $\phi_{rs}=0$ if  $\mbox{tr}\! \left[ \rho\, (P_r^A A_0) \!\otimes\! (P_s^B B_0) \right]$ is positive, and $\phi_{rs}=\pi$ if it is negative. Because the expected value \w{expected} is real, it can be written as
\be
  \sum_{r,s=0,1}  \cos(\phi_{rs} + ra\alpha + sb\beta) \left| \mbox{tr}\! \left[ \rho\, (P_r^A A_0) \!\otimes\! (P_s^B B_0) \right] \right|. \nonumber
\ee
It is easy to see in the last expression, that a generic vector $\langle A_a B_b \rangle$ can always be written as a convex combination of the vectors 
\be
  \{\langle A_a B_b \rangle = \cos(\phi + a\alpha + b\beta);\ \forall \phi, \alpha, \beta \} \ ,
\label{bombai} \ee
because the positive numbers $\left| \mbox{tr}\! \left[ \rho\, (P_r^A A_0) \!\otimes\! (P_s^B B_0) \right] \right|$ sum up to no more than one, and the null vector belongs to (\ref{bombai}). The relabeling of the angles 
\bea
  &\phi_1 =& \frac{\pi}{2} - \phi \\
  &\phi_2 =& \frac{\pi}{2} + \phi + \beta \nonumber \\
  &\phi_3 =& \frac{\pi}{2} + \phi + \alpha \nonumber 
\eea
transforms (\ref{bombai}) in (\ref{generators}), hence, both sets are the same. Second step: it can be checked that, all vectors in (\ref{bombai}) can be obtained by measuring the two-qubit maximally entangled state
\be
  \ket{\Phi_{AB}} = \frac{1}{\sqrt{2}} \left( \ket{0_A}\!\ket{0_B} + e^{i\phi} \ket{1_A}\!\ket{1_{B}} \right), 
\ee
with the observables
\bea
  &A_a& = e^{ia\alpha}\ket{0_A}\!\bra{1_A} + e^{-ia\alpha}\ket{1_A}\!\bra{0_A} \\
  &B_b& = e^{ib\beta}\ket{0_B}\!\bra{1_B} + e^{-ib\beta}\ket{1_B}\!\bra{0_B} \ .
\eea
As we have said before, this concludes the proof of Lemma 4. $\Box$

\bigskip
{\bf Lemma 5:} $G$ has the same symmetries than $Q$.

The exact meaning of this statement is that, the set $G$ remains invariant under any permutation of the coordinates and under change of sign of any two coordinates. A way to show this fact is writing the definition of $G$ \w{generators} in a different way:
\be
  \begin{pmatrix} \sin\phi_1 &\, \sin\phi_2 &\, \sin\phi_3 &\, \sin\phi_4 \end{pmatrix}
\label{memo} \ee
where the constraint
\be
  \phi_1 +\phi_2 +\phi_3 +\phi_4 = \mbox{ multiple of } 2\pi 
\label{rel} \ee
must hold. Now, the invariance under permutations becomes manifest. Changing the sign of two coordinates, say $i$ and $j$, is equivalent to the transformation
\bea
  \phi_i \rightarrow \phi_i + \pi \nonumber \\
  \phi_j \rightarrow \phi_j + \pi 
\eea
which keeps the constraint \w{rel} satisfied. $\Box$

\bigskip
{\bf Lemma 6:} $G$ is contained in $Q$.

The result of Lemma 5 allows us prove this lemma considering only s-ordered vectors. Then, we just have to check that all s-ordered points in \w{generators} satisfy \w{q2}, that is
\be
  f(\phi_1)+f(\phi_2)+f(\phi_3)+f(\phi_1+\phi_2+\phi_3) \leq \pi \ ,
\label{laden} \ee
where $f(\phi)=\a\!(\sin \phi)$. This function is continuous, periodic and has constant slope ($\pm 1$) within intervals of length $\pi$. The points at which $f(\phi)$ changes its slope are
\be
  \frac{\pi}{2} + \mbox{ multiple of } \pi
\label{ext} \ee
In what follows it is found that the maximum of 
\be
  f(\nu_1)+f(\nu_2)+f(\nu_3)+f(\nu_4)
\label{bin} \ee
constrained to
\be
  \nu_1 +\nu_2 +\nu_3 -\nu_4 =0
\label{cons} \ee
is $\pi$, which proves this lemma. Assume that the maximum is attained for some values of $\nu_1$, $\nu_2$, $\nu_3$ and $\nu_4$. If one $\nu_i$ is not of the form \w{ext}, equation \w{cons} implies that there is an other $\nu_j$ which is neither of the form \w{ext}. We can change the value of these two variables inside the range where $f(\nu_i)$ and $f(\nu_j)$ keep their slope constant, and equation \w{cons} holds. This operation should not increase the value of \w{bin}, otherwise the initial point would not be a maximum. This operation should not decrease the value of \w{bin}, because within the region of constant slope it could be also increased. The increase of $\nu_i$ and $\nu_j$ can be performed until one of them reaches a value of the form \w{ext}. Once this operation has been performed, if there is still an other $\nu_k$ not being like \w{ext}, the previous procedure can be repeated. All this implies that, there is a point with coordinates of the form \w{ext}, satisfying \w{cons}, which makes \w{bin} achieve its maximum value. There is only one s-ordered point with coordinates like \w{ext}: $\nu_1=\nu_2=\nu_3=\pi/2$ and $\nu_4=-\pi/2$, for which the function \w{bin} gives $\pi$. $\Box$ 

\bigskip   
{\bf Proof of the Theorem:} First, let us prove that $\c G \subseteq Q$. The fact that $G \subseteq Q$ (Lemma 6) together with the convexity of $Q$ (Lemma 1) imply that $\c G \subseteq Q$. Second, let us prove that $Q \subseteq \c G$. To accomplish this it suffices to show that $F \subseteq G$, because this implies that $\c F \subseteq \c G$, and Lemma 2 say $Q \subseteq \c F$. Then, the only thing that has to be done in order to prove that $\c G = Q$ is to show that $F \subseteq G$. Considering only s-ordered vectors it is enough to show that all points saturating \w{q2} ---which are the ones for which \w{eq} holds--- are of the form \w{generators}. This is straightforward after identifying $\phi_1=\gamma_1$, $\phi_2=\gamma_2$, $\phi_3=\gamma_3$ and $\phi_4=-\gamma_4$. Finally, recalling that all correlation vectors obtainable within QT are $\c G$ (Lemma 4), the proof of the theorem is finished. $\Box$ 

\bigskip
{\bf Final Remarks.} There are many different probability distributions giving the same correlator. This can be seen in the expression relating both:
\bea
  \langle A B \rangle = &p(A=1,B=1) + p(A=-1,B=-1)& \nonumber \\
  -&p(A=1,B=-1) - p(A=-1,B=1)&
\eea
It was proven in \cite{Fine} that if a vector of correlators $\x$ satisfies \w{chsh} then, all probability distributions associated to $\x$ are achievable with LVTs. What has been proven in this paper concerning QT is slightly weaker than this. The fact that a vector of correlators $\x$ satisfies \w{nonlinear} only implies that there exists at least one probability distribution for each correlator in $\x$ predictable by QT.

All the results obtained in this work could be generalized to $n$ parties. In \cite{Werner} there were found the generalizations of $C$ and $Q$, let us call them $C_n$ and $Q_n$. While $C_n$ is a polytope, $Q_n$ is a complicated convex set characterized in terms of a convex hull like \w{generators}. Consider now the non-linear map
\be
  \mu(\x)_i = \frac{2}{\pi} \,\a x_i ,
\label{mu} \ee
which is bijective in the set of correlation vectors. This map has the nice property that transforms a complicated convex set into a simple polytope: 
\be 
  \mu(Q_2)=C_2 
\label{inch} \ee
The set of generators of $Q_n$ ---being \w{generators} for $n=2$--- found in \cite{Werner} has the same structure for any $n$: each component is the sinus of a linear function of $n+1$ variables. This suggests that for any $n$, after performing the map $\mu$, the set of quantum correlation vectors becomes a polytope. Even more, it could be that it becomes the LVTs polytope $C_n$, as happens for $n=2$ \w{inch}. It can be seen that the last is not true recalling that, for $n=3$ there are exact contradictions like the one for the GHZ state \cite{GHZ}:
\bea
  &\langle A_0 B_0 C_1 \rangle& = \langle A_0 B_1 C_0 \rangle = \langle A_1 B_0 C_0 \rangle = 1 \nonumber \\
  &\langle A_1 B_1 C_1 \rangle& =-1 
\label{ghz} \eea
The fact that the map $\mu$ leave invariant the components equal to $\pm 1$ implies that the image of \w{ghz} according to $\mu$ still contains the contradiction, therefore, it cannot belong to the LVTs polytope. When $n>3$ three of the parties can share a GHZ state and obtain again a contradiction. Thus, $\mu(Q_n)$ is not equal to $C_n$ when $n>2$, nevertheless, it is quite probable that $\mu(Q_n)$ is a polytope. A polytope is always characterizable by a set of linear inequalities, that in the worst case can be found numerically. In this case, the sets $Q_n$ could be characterized in a simple way.

As a final application let us mention that, when a vector of correlators \w{correlations} cannot be obtained measuring entangled states, communication is necessary. The result of this work can be used to compute the minimal average communication sufficient to get any vector of correlators when entanglement is available.

\bigskip
Finally, it worths mentioning the nice resemblance of the equations defining $C$ and $Q$, which are related by the map $\mu$ \w{mu}. The first set in FIG. \ref{fig} is the image of the second one according to $\mu$. Then, one can interprete these figures as pictorial three-dimensional versions of $C$ and $Q$ respectively.

\bigskip
{\bf Acknowledgements:} The author is grateful to A. Ac\'{\i}n, J. I. Latorre, A. Prats for helpful comments and suggestions, and G. Vidal for noticing the appearance of s-order in this problem. This work is financially supported by the projects MCYT FPA2001-3598, GC2001SGR-00065 and the grant 2002FI-00373 UB.

\end{document}